# Planetary Science Goals for the Spitzer Warm Era

C.M. Lisse[*], M.V. Sykes[†], D. Trilling[**], J. Emery[‡], Y. Fernandez[§], H.B. Hammel[¶], B. Bhattacharya[‖], E. Ryan[§§] and J. Stansberry[**]

[*]*Planetary Exploration Group, Space Department, Johns Hopkins University, Applied Physics Laboratory, 11100 Johns Hopkins Rd., Laurel, MD 20723, USA*
[†]*Planetary Science Institute, 1700 E. Fort Lowell Rd, Suite 106, Tucson, AZ 85719, USA*
[**] *Steward Observatory, University of Arizona, 933 N. Cherry Ave, Tucson, AZ 85721, USA*
[‡] *SETI Institute / NASA Ames Research Center, Mail Stop 245-6, Moffett Field, CA 94035, USA*
[§] *Department of Physics, University of Central Florida, 4000 Central Florida Blvd, Orlando, FL, 32816-2385, USA*
[¶] *Space Science Institute, 4750 Walnut Street, Suite 205, Boulder, CO 80303, USA*
[‖] *Spitzer Science Center, California Institute of Technology, Pasadena, CA 91125, USA*
[§§]*Department of Astronomy, University of Minnesota, 116 Church St SE, Minneapolis, MN 55455, USA*

**Abstract.** The overarching goal of planetary astronomy is to deduce how the present collection of objects found in our Solar System were formed from the original material present in the proto-solar nebula. As over two hundred exo-planetary systems are now known, and multitudes more are expected, the Solar System represents the closest and best system which we can study, and the only one in which we can clearly resolve individual bodies other than planets. In this White Paper we demonstrate how to use Spitzer Space Telescope InfraRed Array Camera Channels 1 and 2 (3.6 and 4.5 μm) imaging photometry with large dedicated surveys to advance our knowledge of Solar System formation and evolution. There are a number of vital, key projects to be pursued using dedicated large programs that have not been pursued during the five years of Spitzer cold operations. We present a number of the largest and most important projects here; more will certainly be proposed once the warm era has begun, including important observations of newly discovered objects.

## 1. INTRODUCTION AND OVERVIEW

Condensable elements produced by stellar nucleosynthesis form icy volatile and refractory dust particles that eventually find their way into young stars and planetary systems. Dust is known to orbit a growing number of main sequence stars (Beta Pic, Vega, Epsilon Eri, and HD100546 are famous examples).

The dust in these stars is distinguished by having a lifetime that is short compared to the main-sequence lifetime of the central stars. Therefore, the dust cannot be primordial but must have been recently produced, perhaps by collisions among parent bodies in an unseen analog to the Kuiper Belt (KB) or asteroid belt, or by sublimation of comets. Much of the particulate material in such disks is cool, residing in the equivalent of the KB region. Strong similarities are seen between the spectra of dusty material emitted by comets, and of the dust in exo-systems (Fig. 1; Lisse *et al.* [1], [2]). The clear implication is that small bodies must exist in other planetary systems.

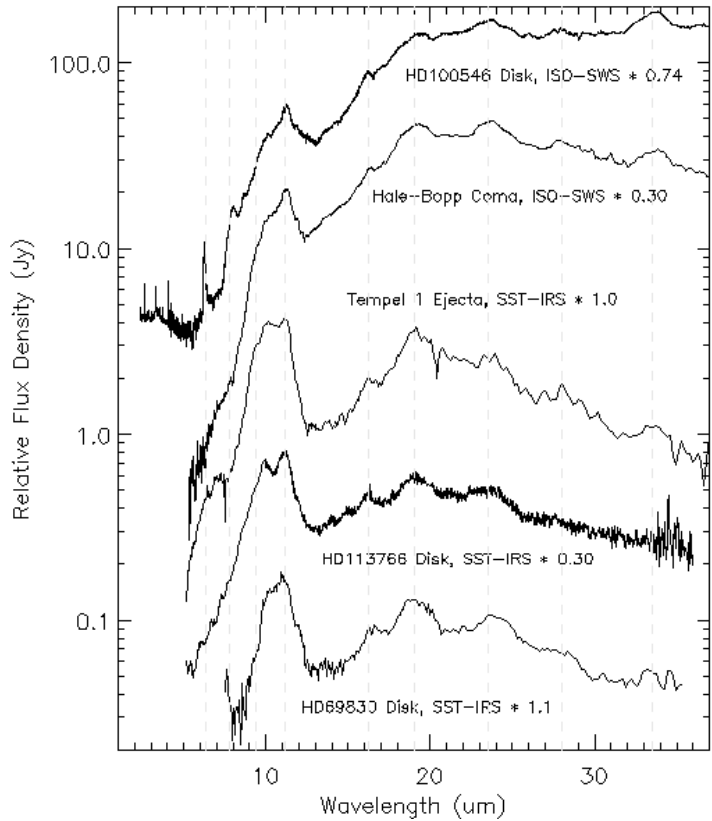

**Figure 1.** Similarities of mid-IR spectra of Solar System comets and dust found around other stars. Mid-IR spectra of the YSO HD100546, the young terrestrial planet building system HD113766, the mature solar-system like HD69830 system, and the comets C/1995 O1 (Hale-Bopp) and 9P/Tempel 1, showing the gross similarities and differences in the emission for the sources. Most of the obvious differences between the HD100546 and Hale-Bopp and the Tempel 1, HD113766, and HD69830 spectra are due to temperature - the latter three sources are much warmer on the average, and thus have relatively more emission in the 5-10 μm region. The other obvious difference is the very strong PAH emissions for HD100546 at 2 - 9 μm, and the superabundance of olivine in HD69830, as discerned from the pattern of the strong 9 - 12, 16 - 20, and 23 - 25 μm emission features.

The current paradigm of planetary system formation invokes the condensation of refractory materials (this term can be understood roughly as "rocky" or "solid" materials) in the thick proto-solar disk, surrounded by a gaseous envelope. Eventually the solid grains settle into a thin disk, whereupon aggregation into larger bodies begins. Dust grains grow to become pebbles, then boulders, then kilometer-sized and hundred-kilometer-sized planetesimals - comets and asteroids - the building blocks of solid planets. Planet formation has led to the clearing of the sun's disk inside about ~30AU. Many planetesimals in the inner Solar System were incorporated into the present day terrestrial planets, but many persist as 1- to 1000-km asteroids.

Planetesimals that formed in the region of the gas giants were either incorporated into the planets or rapidly ejected. Most escaped the Solar System entirely, but a large population was captured by stellar and galactic perturbations into the Oort cloud. Oort comets may be gravitationally scattered back into the inner Solar System to appear as dynamically new long-period comets, some of which are eventually captured into Halley-family short-period orbits.

The heavyweights of our Solar System are the giant gaseous planets. These formed from the planetesimals, like the terrestrial planets, but were able to also incorporate nebular material directly from the protoplanetary disk into their makeup, and were thus able to retain even the most volatile of molecular species found in the proto-solar nebula, like CO, $CH_4$, H, and He. With more than 200 known Jupiter-to-Neptune mass planets known to be orbiting other stars, this process of giant-planet formation is clearly ubiquitous. Jupiter and its kin hold clues to understand the processes that govern planets in other planetary systems.

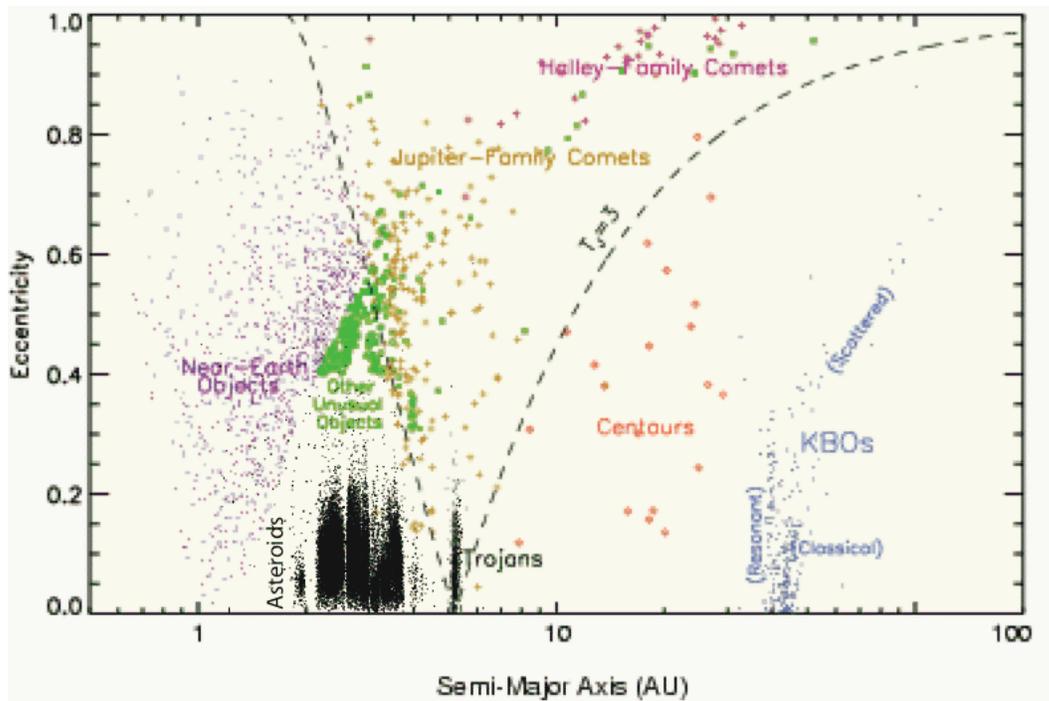

**Figure 2.** The major families of Solar System objects, as revealed by their dynamical structure**.** The Solar System's small bodies are shown plotted as a function of orbital semimajor axis (in AU) and orbital eccentricity. On the scale of this figure, the Sun and planets lie at approximately zero eccentricity. For clarity we reduced the eccentricity value of the asteroids by a factor of 2. The boundaries are not impermeable; for example there are inactivated asteroids in the nominal Jupiter-family comet region, and there are active comets in the near-Earth object region. This overlap hints at the evolutionary relationship among the groupings.

In the cool outer disk (beyond Jupiter's orbit), the planetesimals incorporated frozen volatiles as well as refractory material. The trans-Neptunian region is occupied by numerous bodies with sizes up to several thousand km (Pluto and other dwarf planets). Collisions in this population generate fragments, some of which were scattered inwards by dynamical chaos to become intra-planet wanders. These wanderers have short dynamical lifetimes, and are ejected from the Solar System entirely through gravitational interactions, or scatter further inward, where they begin to sublimate and are re-classified as first Centaurs in the giant planet region, and then Jupiter-family short-period comets in the inner Solar System. Comets that lose all their volatiles from prolonged solar heating appear as asteroids having unusual, comet-like orbits. A number of these dead comets are known among near-Earth asteroids, and are also prime targets for the Warm Era (Fig. 2).

In this scenario, there is a clear evolutionary flow-down of protoplanetary material, through planetesimals (KBOs, Centaurs, comets and asteroids) to planets. Many key questions in the details of this scenario of planetary system formation -- for our system, and others -- remain unsolved. We briefly list here six of the most relevant questions addressable in the Warm Spitzer Era (in rough priority order of importance), and describe each more completely below.

(1) **What is the dynamical history of the Solar System, and what role has giant-planet migration played?** There is substantial theoretical work suggesting that giant planets (in our Solar System and in others) undergo significant radial migration before the system settles into its final configuration, but there is still little direct evidence for this hypothesis. A Warm Spitzer can provide observational tests of specific predictions made by those theories by observing various small bodies in the Solar System. The results of those tests have strong implications for the dynamical evolution of our Solar System. [Trojans, Outer Main Belt Asteroids (MBAs), Kuiper Belt Objects (KBOs)]

(2) **What is the distribution of water and organic components in the Solar System?** The dynamical evolution of our Solar System has direct implications for the habitability of the Earth. It has been proposed that water and organic material was delivered to the Earth by impacting comets and asteroids. With a Warm Spitzer, we can map the distributions of water and organic material in the asteroid belt and constrain the consequent flux rate of this life-giving material to the early Earth. [Main Belt Asteroids]

(3) **What are the physical properties of potential Earth-impacting asteroids?** Survival on the Earth might be an "easy come, easy go" game -- an impact 65 million years ago eliminated the dinosaurs and allowed mammals to rise to prominence. Equally, there are thousands of asteroids known in near-Earth space. With a Warm Spitzer, we can understand the physical properties of these bodies and help evaluate the potential impact hazard onto the Earth. This information will potentially aid preparations for mitigating the impact hazard through direct manipulation of those asteroids. [Near-Earth Asteroids]

(4) **What is the nature of the refractory and carbonaceous material incorporated into comets from the proto-solar nebula?** Carbon is critical to the beginnings and evolution of life, as well as being an important component of the Earth's crust. Comets -- as leftover planetesimals from the era of Solar System formation -- contain important details of the mechanism of aggregation of ISM dust and gas into planets. The major reservoir of carbon in comets is in the gaseous volatile organic and $CO/CO_2$ ices, not in the refractory state, and is poorly studied. [Comets]

(5) **Are the structures we see in mature exo-disks explainable by the erosional and collisional processes supporting the Solar System zodiacal cloud?** For example, the major source of mass loss from active comets is in large (supermicron), dark trail particles that are an important source for our Solar System's interplanetary dust cloud. Can the model timescales for these processes (10 - $10^6$ yrs for cometary emission processes, $10^6$ - $10^7$ yrs for asteroid fragmentation events, $10^5$ yrs for dust-dust collisions, $10^3$ - $10^6$ yrs for PR drag, < $10^6$ years for radiation pressure blowout) be reconciled with zody cloud studies? [Zodiacal Cloud]

(6) **Is the major energy source driving the dynamics of the icy giant planets Uranus and Neptune gravitational contraction, insolation, or dynamical friction?** Answering this question is naturally crucial for understanding the energy budgets of exoplanet atmospheres. Current observations of Uranus suggest that seasonal insolation changes are more significant than predicted by models, and that Neptune's atmosphere may be more sensitive to insolation than expected. By extending observations of Uranus and Neptune over the five years of the Warm Era -- through major events in the seasonal cycles of the two planets -- we can determine which effects are paramount. [Ice Giants]

In sum, Solar System astronomy is unique among all fields of astronomy because of our ground truth: we can and do visit many of these primordial bodies with spacecraft missions to asteroids, comets, and planets, and we obtain telescope data of a quality and scope not possible for any other planetary system. (The ground truth also comes to us naturally, in the form of meteorites!) The Spitzer observations described below serve as a bridge between our rare but detailed ground truth data and understanding the formation and evolution of planetary systems throughout the galaxy.

## 2. SMALL OUTER SOLAR SYSTEM BODIES: TNOS, CENTAURS, JOVIAN TROJANS, AND OUTER MAIN BELT ASTEROIDS

Spitzer warm era observations are perfectly suited to determination of surface compositions of Trans-Neptunian Objects (TNOs), Centaurs, and Trojan and outer Main Belt asteroids through measurements of broadband reflected fluxes with IRAC. The value of IRAC is that its measurements of reflectance will 1) provide a far more sensitive search for ices than is possible at shorter wavelengths and 2) readily distinguish between candidates for the poorly understood "dark material," which is also not possible at shorter wavelengths. These capabilities critically address several longstanding questions in planetary science related to the nature of the nearly ubiquitous "dark material" and the distribution of volatiles, and more broadly address the following key question: What is the dynamical (and chemical) history of the Solar System, and what role has giant planet migration played?

### 2.1 Dynamical Evolution of the Outer Solar System

Two general views of the origin and dynamical evolution of the Solar System are currently in circulation. One, built up from a wealth of data and modeling in the mid to latter part of the last century, posits a relatively dynamically quiescent system after accretion. Starting in the 1990s, contemporaneous discoveries (and dynamical characterization) of Kuiper Belt objects in the Solar System and giant planets in exotic orbits around other stars have lead to increased consideration of giant planet migration and the dynamical eruptions such migrations induce. A second leading hypothesis for the Solar System has emerged in which all four giant planets have undergone migration, with severe implications for the minor body populations. In either scenario, the groups of primitive bodies we propose to observe are inherently interesting because of the information they hold on the compositional make-up and chemical evolution in critical regions of the nebula as well as the current distributions of astrobiologically and cosmochemically important materials (i.e., $H_2O$ and organics). Furthermore, their importance increases significantly by the fact that their origins, and therefore compositions, provide the opportunity to distinguish between the two hypotheses for the dynamical evolution of the Solar System.

The Jupiter Trojans are particularly relevant for distinguishing these hypotheses. Orbiting the Sun at ~5.2 AU (trapped in Jupiter's stable Lagrange points), Trojans have traditionally been thought to have formed near their present location and to have been trapped in the Lagrange points later, perhaps as Jupiter rapidly grew (Marzari and Scholl [3]). In this case, they would represent material from the middle part of the solar nebula where ice first began to condense, a region not sampled by any other class of primitive body. Recently, however, Morbidelli *et al.* [4] proposed a scenario in which the Trojans formed in the Kuiper Belt, were scattered inward and were trapped in the Lagrange points as Jupiter and Saturn crossed their 2:1 mean-motion resonance. In this case, the Trojan asteroids would represent the most readily accessible depository of Kuiper belt material. But more importantly, *the Trojans offer a critical test of the planetary migration model of Morbidelli et al., which has implications for not only the Trojans, but for the dynamical evolution of the Kuiper belt and the Solar System as a whole.*

Trans-Neptunian objects (TNOs; perihelia beyond Neptune's orbit) also hold a wealth of information about the dynamical evolution of the Solar System. Several dynamical sub-classification systems have been proposed within this group, all of which include a cold classical population that has not suffered much scattering, resonant objects that are trapped in mean-motion resonances with Neptune (e.g., Pluto), and a scattered population. Evolution models call upon various mechanisms to explain different aspects of the current dynamical state, including migration of all the giant planets, several now lost (or hidden at very large distances) Neptune-class planetesimals, galactic tides, and close encounters with another star, among others (e.g., Chiang *et al.* [5] and references therein). Physical studies offer the best means to discriminate among these mechanisms. Currently, the only correlation between physical and dynamical properties is that the cold classical objects have systematically redder colors than other objects. Otherwise, the full range of spectral types (red, neutral, featureless, volatile-bearing) is equally well represented in the dynamical classes. As knowledge of the compositional types improves (by observations like those suggested here), it should be possible to identify source regions for the different compositional types. Then evolution can be run in reverse, and the different proposed dynamical mechanisms tested. Indeed, results from a program to observe a small number of TNOs using IRAC (Spitzer program 20769) hint at groupings not previously seen (see Fig. 3). We expect that insights from this new result will lead to eventual unraveling of the evolutionary history of the Kuiper Belt as discussed above, but observations of a much larger sample of objects are needed to flesh out and understand these groupings.

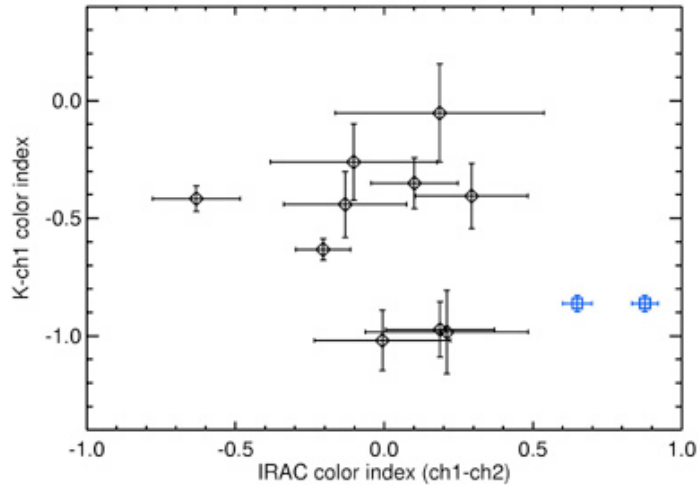

**Figure 3.** IRAC data of TNOs from program 20769 shows a bimodal distribution. The two IRAC observations made of each object were averaged together for this plot. The blue squares on the right of the plot are Pluto data at two different longitudes.

## 2.2 Composition of Dark Material in the Outer Solar System

Most TNOs, Centaurs, Trojans, and outer-belt asteroids have low visible albedos. The nature of this "dark material" that lowers the albedos is almost completely unknown. The reddest Centaurs and TNOs require organics to explain the steep spectral slopes in the visible and NIR (Cruikshank and Dalle Ore [6]). The analog species that have been put forth a) all have very strong absorptions in the wavelength range of the IRAC bands, and b) have spectra that often differ widely from one another in the region covered by the IRAC bands (compare the green dotted and red solid curves in Fig. 4). Surface compositions of the less red objects are even more uncertain. The neutral to moderately red spectra can be explained equally well with silicates or a variety of organics (e.g., Cruikshank et al. [7], Emery and Brown [8]). Moderately red-sloped objects (including KBOs, Centaurs, and asteroids) are common in the Solar System. Distinguishing their surface compositions is crucial for understanding the distributions of organics, which is an issue that bears significantly on ideas of the chemistry and processes in the solar nebula. Silicates and organics have very different spectral behavior at $\lambda > 3$ μm.

IRAC observations therefore offer the best means to constrain the class of material that causes both the ultra red spectral slopes and the more moderate slopes. We note that the IRAC data of the ultra-red Centaur Pholus and the scattered disk object Sedna (Fig. 4a), which were observed by IRAC as part of program 20769, are both best fit with nitrogen-rich tholins as opposed to more oxygen- or C-H rich organics. Similarly, IRAC data of the moderately red Centaur Asbolus suggest N-rich tholins rather than either C-H rich organics or silicates (Fig. 4b). Since there is such a wide range of possible compositions for these low-abedo, red objects, it is particularly important to observe a large sample and look for correlations between dynamical and compositional properties.

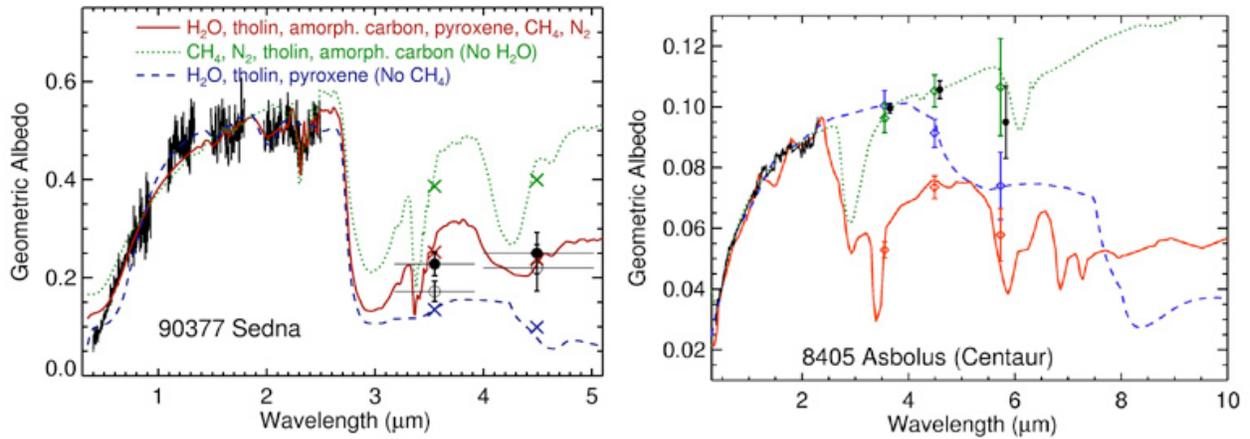

**Figure 4**. (a; left) IRAC data for the TNO Sedna (filled and open circles) along with the vis-NIR spectrum and three spectral models. (b; right) IRAC data of the Centaur 8405 Asbolus (black circles) compared with three compositional models that all match the shorter wavelength data.

## 2.3 Volatile Inventory of the Outer Solar System

$H_2O$ ice has been detected on several TNOs and Centaurs through absorptions at 1.5 and 2.0 μm, but others do not exhibit these bands. Trojan and outer-belt asteroids may have $H_2O$ ice in their interiors, but none has yet been detected. If these asteroids originated in the same region as TNOs, they should have icy interiors, and we would expect impacts to uncover some of these ices, particularly among dynamical family members. Because $H_2O$ contains very strong absorptions at $\lambda > 2.7$ μm, the 3.6 μm IRAC channel provides a sensitive test. IRAC data points are also sensitive to the presence of other ices, such as $CH_4$ and $CH_3OH$ (see below). The greater sensitivity to ices that IRAC offers will more robustly identify correlations between the presence of volatiles and any other dynamical or physical properties, if they exist.

Ten of the 11 KBOs observed so far from program 20769 show strong absorption at $\lambda > 3\ \mu m$ in the IRAC data (Emery *et al.* [9]). This includes the first detection of ices on four objects. In addition, Emery *et al.* [10] have used IRAC data to identify the first reported detection of $H_2O$ on Sedna (Fig. 4a), along with confirmation of the $CH_4$ reported from shorter wavelength data (Barucci *et al.* [11]).

## 2.4 Pluto and Kin

The dwarf planets Pluto, Eris, Sedna, 2005 FY9, and 2003 EL61 are particularly interesting because of their potential for atmospheric and geologic activity. Near-IR spectra reveal the presence of $CH_4$ on all of these objects except 2003 EL61. Some spectra also suggest the presence of $N_2$ ice on Eris, and Pluto has abundant $N_2$.

Photolysis and cosmic ray radiolysis are expected to quickly convert $CH_4$ to higher-order carbon molecules, with resultant darkening of surface layers. However, Spitzer MIPS radiometry and HST imaging show that the albedos of these objects are quite high. This is probably the result of seasonal transport of the $CH_4$, which will segregate the volatile ice from the photolytic products. The seasonally active layer should result in changes of the albedo patterns on the surfaces of these objects, and any others which may be discovered over the next few years. The geographic distribution of the volatile ices directly influences their temperature, and thereby the pressure of their vapor in the atmosphere. As the ices are transported across the surface by sublimation, both the appearance of the surface and the atmospheric pressure can be expected to undergo seasonal changes.

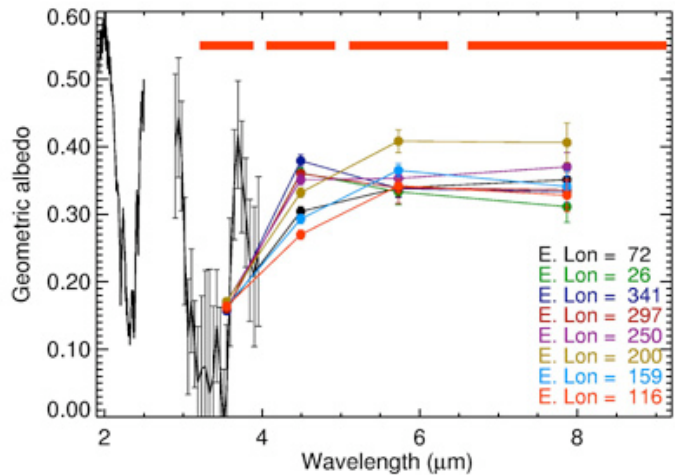

**Figure 5**. IRAC measurements of Pluto from GTO program 70 along with a ground-based NIR (1.9 - 4.0 µm) spectrum. The colored filled circles are IRAC data at eight different Pluto longitudes.

Detailed IRAC observations of these kinds of objects are an example of a small warm-mission program that could reveal very interesting seasonal behavior, and also significantly deepen our understanding of the compositions of their surfaces. For example, the IRAC data for Sedna (Fig 4a) clearly provide a unique test for constraining the composition in ways that cannot be achieved using current ground-based capabilities. IRAC data for Pluto (Fig. 5), to which Eris and 2005 FY9 are best compared, revealed the expected $CH_4$ absorption in the 3.6 *µm* channel. The 4.5 *µm* reflectance, however, has a rotational variation that is distinct from the other three bands, the visible albedo, or any of the NIR spectral features (Grundy and Buie [12]). This manifests itself as two distinct spectral units on Pluto in the IRAC channels. This may be due to organic materials or previously undetected ices. Rotationally resolved IRAC light curves for other objects in this class could also reveal unique compositional and geographic information that cannot be obtained using any other existing capability.

# 3. MAIN BELT ASTEROIDS

There are more than 350,000 known Main Belt asteroids. Many of them contain volatiles -- most notably water, in various forms -- and also organic material, both of which are retained from the era of planet formation 4.5 billion years ago. The traditional paradigm of Solar System formation suggests a condensation sequence from inner to outer Solar System in which high temperature silicates condensed in the inner Main Belt (E-type asteroids), followed by moderate temperature silicates (S-type asteroids) in the inner to mid Main Belt, low temperature silicates with a small fraction of organics in the mid-outer Main Belt (C-type asteroids), and increasing organic material further out (P- and D-type asteroids). Most mid-belt C-type asteroids contain hydrated silicate spectral features, which have been attributed to a post-accretion heating event - possibly magnetic induction heating - whose strength was a function of heliocentric distance. A new model proposes large-scale dynamical upheaval throughout the outer Solar System, which may have affected the mid to outer Main Belt. In this new scenario, the outer Main Belt asteroids in particular may have formed much further out in the Solar System. Furthermore, important non-equilibrium chemical pathways have been identified that can account for the formation of hydrated silicates in the nebula rather than on asteroid parent bodies themselves. Whereas in some cases meteorite petrology strongly indicates in situ formation of the hydrated silicates, in other cases formation on the body is not certain. Additionally, there has to date been no direct evidence (e.g. spectroscopic) to support the hypothesis that the low albedos and red spectral slopes are due to organic materials. Mapping these components in the main belt allows us to understand the distribution of volatiles and organic material in the early, nascent Solar System.

We therefore ask the following key critical question: *What is the distribution of water and organic compounds in the Solar System?*

Water ice, hydrated minerals, and organic materials all exhibit very strong spectral signatures in the 3 to 5 $\mu m$ range. Spitzer IRAC observations of outer main belt asteroids, with location beyond 3 AU, provide a clear measure of the thermal emission from the asteroid surface, and therefore offer the means to perform very sensitive, systematic searches for these diagnostic materials on a large sample. The two IRAC bands are not sufficient to determine mineralogy, or in most cases even to distinguish among $H_2O$, hydrated minerals, and organic materials. However, identification and classification of absorptions using IRAC provides an important search that can be used to refine future, more detailed spectral studies with, for example, JWST.

We tested the feasibility of such a project using spectral models. A model fit that used hydrated minerals and organic materials but ***no*** $H_2O$ (solid black dot in Fig. 6) was computed for a typical outer-belt D-type asteroid, then varying amounts of the absorbing material were added to investigate the effects on the K-3.6μm and 3.6μm-4.5μm color ratios. S/N ~10 will be sufficient to detect even just a few percent of these materials on the surface. Although only $H_2O$ and hydrated minerals are shown on this plot, the effects are similar with macromolecular organics (e.g. tholins). Measurements of K-3.6μm and 3.6μm -4.5μm colors (using 2MASS with Warm Spitzer) can be used to pull out those compositions.

There are ~100,000 asteroids with orbital semi-major axes greater than 3 AU, of which ~10,000 are in the 2MASS database. To meet out science goals, a sufficient number of target asteroids can easily be selected from the 2MASS subset, choosing objects with favorable apparitions and a range of sizes, as appropriate.

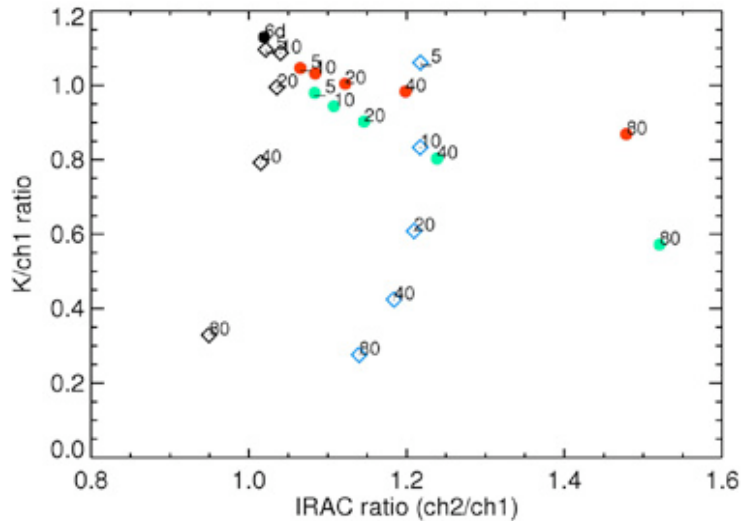

**FIGURE 6.** Band ratios for various fractions of $H_2O$ ice (black diamonds - 100 μm grains, blue diamonds - 12.5 μm grains) and hydrated minerals (red filled circles - chlorite, green filled circles - serpentine). The black filled circle represents the volatile-free spectrum. The numbers adjacent to each symbol indicate the percentage of the absorber that is included in the mixture.

## 4. NEAR EARTH OBJECTS

65 million years ago, a 10 km body hit the Earth and caused the extinction of the dinosaurs. Asteroids and comets are known to hit the Earth periodically, with a frequency that is inversely proportional to their size; discussion of the next impact onto the Earth is a question of "if" not "when." The U.S. Congress has established a requirement that 90% of the potential Earth-impacting asteroids larger than 140 meters be identified by the year 2020. In this way, potential hazards can be identified, and mitigation plans defined, if necessary. These potentially threatening asteroids have been named Potentially Hazardous Asteroids (PHAs), and generally fall within the dynamical category of Near Earth Object (NEOs). PHAs are defined as bodies larger than 140 meters that will pass within 0.05 AU of the Earth's orbit, and NEOs are all bodies whose orbits pass within a few tenths of an AU of the Earth's orbit. As of this writing, there are around 4000 NEOs, and 850 PHAs. The Pan-STARRS program is likely to increase the number of known NEOs to ~10,000 by 2013.

These bodies are of specific interest to humans. Characterizing potential threats to life on Earth carries an almost unimaginably high value. Additionally, these near-Earth objects are the most easily reached by spacecraft, enabling our exploration of the Solar System. These nearest neighbors can be explored with Spitzer during the Warm Era.

Consequently, we ask the following key question: *What are the physical properties of potential Earth-impacting asteroids?*

## 4.1 NEO Sizes and Densities

NEOs have relatively hot surface temperatures (>250 K). Fluxes in IRAC 3.6µm and 4.5µm will therefore be dominated by thermal flux (Fig. 7). We can derive the sizes and albedos of NEOs by combining thermal measurements with reflected light data (i.e., visible magnitudes). At sizes <1 km, these bodies will be the smallest bodies observed by Spitzer. Deriving the sizes of hundreds of NEOs also will allow us to establish a true size distribution, which reflects the dynamical origin of this population.

Approximately 10% of NEOs are binaries. Using Keplerian dynamics, thermal measurements of NEOs that turn out to be binaries will yield densities, which in turn will suggest compositions and internal strengths. The intrinsic strengths of these bodies will be particularly important in mitigating potential hazards to the Earth -- disrupting a solid body is very different than disrupting an unconsolidated rubble pile.

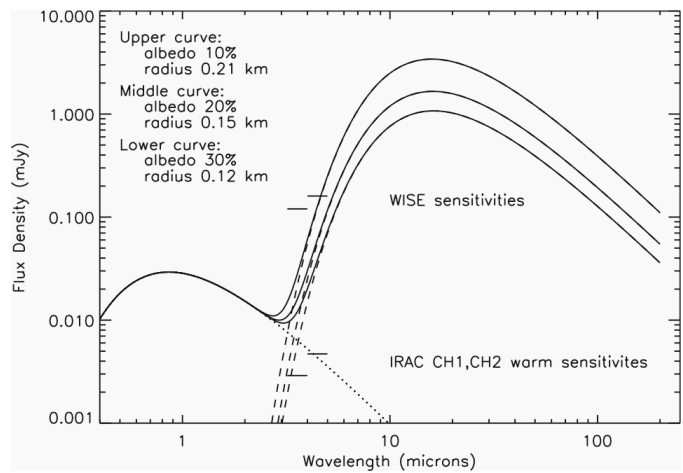

**FIGURE 7.** A predicted spectral energy distribution (SED) for an NEO with heliocentric distance 1.3 AU and Spitzer-centric distance 1.0 AU. Three different albedos are shown (and their corresponding diameters). The lower horizontal lines are Warm Spitzer sensitivity limits, 5σ, 500 seconds (high background). The upper horizontal lines are WISE sensitivities for the ecliptic pole, 5σ, in sky survey mode.

## 4.2 Regoliths and Thermal Inertia

Most airless bodies in the Solar System are covered to some degree with regolith, layers of pulverized rock that are produced over time by collisions with both large and small particles. It has recently been suggested, based on indirect evidence, that bodies smaller than ~5 km may be nearly devoid of regolith (Binzel *et al.* [13], Cheng [14]). This prediction affects the most numerous population of objects in the Solar System (sub-km asteroids) and our nearest neighbors in the Solar System.

The amount of regolith on a surface is reflected in the body's thermal inertia -- the degree to which the body is in instantaneous thermal equilibrium with solar radiation.

With thermal measurements of a sufficiently large number of hundreds of NEOs, the average thermal inertia can be derived, and therefore the typical regolith amount on these small bodies. Understanding the regolith properties of these small bodies will help us understand their dynamical ages in near-Earth space, with consequent implications for the flux of potential bodies into and across the Earth's orbit.

Asteroids that have high thermal inertias emit significant thermal flux on their night sides. In these cases, asteroid orbits can change through the Yarkovsky effect, in which anisotropic radiation of thermal inertia produces a force through which semi-major axes can change. Consequently, understanding - and potentially mitigating - the Earth impact hazard requires detailed analysis of the thermal inertia and consequent Yarkovsky effect on NEOs.

### 4.3 Origins and Compositions

The majority of NEOs are believed to be S class asteroids, a rocky and relatively volatile poor asteroid type. The majority of main belt asteroids are C type asteroids, more volatile and organic rich. The source and dynamical path and evolution of main belt asteroids to NEOs are not well understood. Measuring the albedos of a large number of NEOs will probe the degree to which there is a C class "tail" of NEOs, with implications for the compositions of Earth-crossing (and potentially impacting) asteroids. Although an impact in the present day would have a strongly negative impact on life on Earth, it is likely that volatiles and organic material were brought to a pre-biotic Earth through just this mechanism. Therefore, understanding the compositions of potential Earth impacting bodies has important implications for understanding the past *and* the future of life on Earth.

## 5. COMETS

Comets are planetesimals formed from the circumstellar material after the proto-solar nebula has collapsed to form the nascent Sun and surrounding disk of material. Comets were formed throughout the disk, from the ice line (where water ice is stable) out to its edges. The comets that formed in the giant planet region were either incorporated into the planets or scattered into the Oort cloud as the gas and ice giants cleared their local regions of space during their accretional growth phase, while the comets that formed at the edges remained in place or were scattered as the ice giants migrated outwards, forming the collection of objects now known as the Kuiper Belt. In the current paradigm for the formation and evolution of our Solar System, most of the Short Period comets dwelling mainly in the inner Solar System are collisional fragments of KBOs that have migrated inward, after having been scattered back and forth among the giant planets. The Long Period comets derive from the Oort Cloud, where they have remained until perturbations from passing stars, molecular clouds, and the galactic tide sent them into the inner Solar System. Both families of comets have been subjected to various degrees of evolution---virtually none over the last >4 Gy for those in the Oort Cloud, extensive collisional evolution for those derived from the Kuiper Belt, and extensive sublimation (and possibly fragmentation and collisional) losses for the SP comets.

Current theories predict that cometesimals decoupling from nebular gas drag should have a characteristic radius of 100m, but the known cometary nuclei and KBOs are substantially larger than this. Thus, an additional poorly understood process of aggregation determines the ultimate form of these bodies. Other physical properties may also be affected by the processes of formation, e.g. collisional fragments may vary more in composition, as revealed by the color of their reflected and thermally emitted light. Until there is an accurate observational database on the physical properties of cometary nuclei to guide the development of theory, we will be stymied in our efforts to explain the formation and evolution of the Solar System outside the ice line. This motivates the key question: What clues do the physical and compositional properties of comets give us about planetesimal accretion?

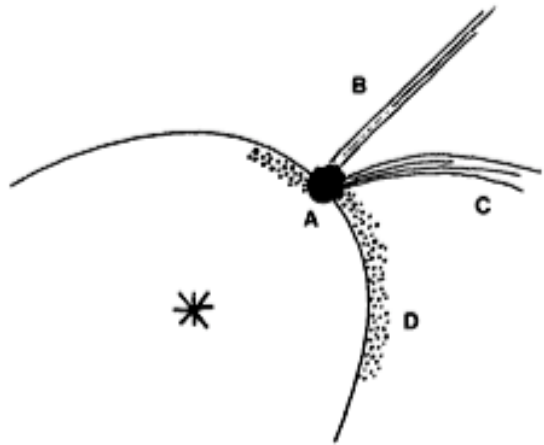

**FIGURE 8.** Schematic of comet morphological structures. (A) Coma, (B) Ion Tail, (C) Dust Tail, (D) Dust Trail (adapted from Sykes and Walker [15]).

## 5.1 Comet Dust Trails

Comets have two fundamental components: refractories and volatiles. The latter are evidenced by gaseous emissions of the coma and ion tail, while the former historically was evidenced by prominent dust tails, consisting primarily of micron-sized particles sensitive to radiation pressure (Fig. 8).

The primary emission of refractory material, by mass, is into the cometary trail, not the oft-studied coma. These dust trail particles were first detected by the Infrared Astronomical Satellite (IRAS) (Eaton et al [16], Sykes et al [17]) and were inferred to be a common property of all short-period comets in a survey of IRAS data (Sykes and Walker [15]). The large trail particles (typically mm-cm in size) are ejected at low velocities relative to the comet orbital speed, which is why they drift away from the comet along orbits close to that of the parent comet, giving rise to the appearance of 'trails' (similar in appearance to fresh airplane contrails). Particles on orbits that come within 3 AU of the Sun (the vast majority of Jupiter family comets) are quickly devolatilized with the result that trail particles allow us to study the refractory component of comets in isolation of volatiles. Trail particles have low-albedos (Sykes *et al.* [18], Sykes *et al.* [19]), making them extremely difficult to study from the ground (e.g., Ishiguro *et al.* [20]).

Spitzer has revolutionized the study of trails using the MIPS 24 $\mu m$ channel (e.g., Fig. 9, Reach et. al [21]); 34 comets were observed, compared with the 8 trails observed by IRAS. Debris trails due to mm-sized or larger particles were found along the orbits of 27 comets; 4 comets had small-particle dust tails and a viewing geometry that made debris trails impossible to distinguish; and only 3 had no debris trail despite favorable observing conditions. There are now 30 Jupiter-family comets with known debris trails. The detection rate is >80%, indicating that debris trails are a generic feature of short-period comets. The mass-loss rate in trail particles is comparable to that inferred from OH production rates and larger than that inferred from visible-light scattering in comae.

Even during its post-cryo phase, Spitzer remains the preeminent facility with pointing capability for studying trails and gaining insights into an important class of Solar System objects, also inferred to exist about other stars. Spitzer allows us to obtain accurate measurements of cometary mass loss rates, details of recent comet emission history, comet dust to gas mass ratios, and the correlation of these properties with dynamical age and history. These studies allow for the probing of the diversity of comet formation locations and source populations. Understanding the cumulative mass loss of comets allows for a better understanding of their role as a source of the zodiacal dust cloud and the question of whether the existence of equivalent exo-clouds depend upon the transport of such bodies from the outer to the inner Solar System or whether such exo-clouds arise from collisional activity among asteroids.

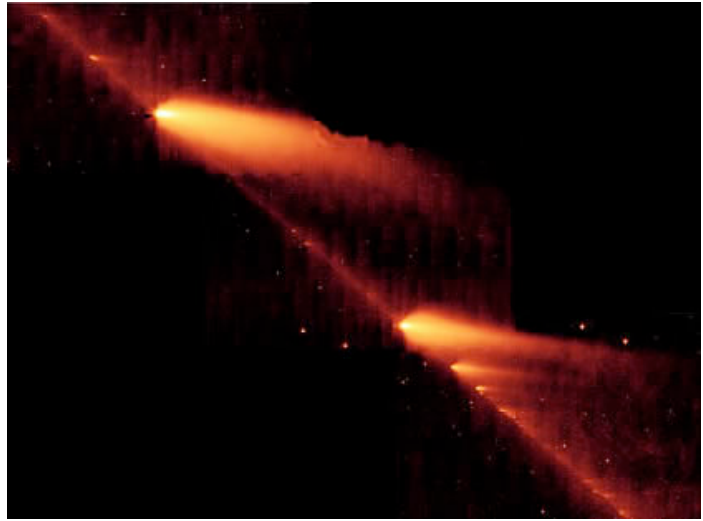

**FIGURE 9.** The P/Schwassmann-Wachmann 3 dust trail and fragmented nucleus form a number of small comets, obtained at 24 µm by Spitzer MIPS (Vaubaillon and Reach [22]). Here the comae are the bright white regions surrounding each fragment, the dust tail the yellowish to reddish extensions directed away from the Sun to the right, and the comet trail is the thin red line running from upper left to lower right, connecting all the fragments. Image courtesy of NASA/JPL-Caltech/W. Reach (SSC/ Caltech).

The Wide-field Infrared Survey Explorer (WISE) is scheduled for launch in 2009. WISE is a survey instrument with a fixed scanning campaign that does not allow for trail observations at optimal viewing geometries and times (within the solar elongation constraints common to both spacecraft). So, Spitzer will be able to use its more flexible pointing capabilities to observe trails at times and geometries unavailable to WISE, as well as conduct follow-on observations of trails observed by WISE. The two systems will be directly complementary.

## 5.2 Carbonaceous Species in Comets

The most important volatile cometary species, water, is often studied in comets using 'hotband' near-IR emission lines that avoid water absorption by the cold Earth's atmosphere. The astrophysically important carbon-bearing molecular species, CO and $CO_2$, are however much harder to observe from the ground and are only well measured in the brightest (most active) of comets. Studies of such comets indicate that these two species are present at the 5 to 20% level versus water in cometary nuclei, making them the second and third most abundant volatile species in comets. CO and $CO_2$ are the major reservoirs for carbon in comets; by comparison, carbon in refractories is roughly only one-fourth as abundant (Lisse *et al.* [2]). We need to detect and measure the dominant carbon volatile species if we are to understand the composition of the planetesimals from which the planets formed.

The budget of CO and $CO_2$ in comets has been unmeasured except for (as stated) the brightest – and therefore most atypical and unusual -- comets. There are almost no detections of these species in any Short Period (inner Solar System) comets, or in any "typical" Long Period comet. It is likely that we are completely missing as much as 25% or more of the volatile content of the majority of comets. Clearly this is a conundrum that needs to be investigated if we are to understand cometary composition. Our hypothesis is that the limiting factor is magnitude; comets are often simply not bright enough for infrared or millimeter spectroscopy to detect CO, and $CO_2$ has no feasible transitions in atmospheric windows. We expect that CO and $CO_2$ are there in the fainter comets, but there can be almost no ground-based measurements to quantify this. Another factor for the lack of detection could be that much of the CO and $CO_2$ in the topmost centimeters of the surface has sublimated away due to surface evolution.

A method of detecting CO and $CO_2$, using IRAC, has been found serendipitously by W. T. Reach and M. S. Kelley (priv. comm.) and provides an unprecedented and unique way to directly test the hypothesis of CO and $CO_2$ abundance in comets. Observations of Comet 2P/Encke in IRAC 3.6μm and 4.5μm indicate a distinct difference in the morphology of the coma that is stronger than would be expected if the coma were simply dust. The shape of the dust coma at 3.6 and 4.5 μm should be virtually identical, since size-dependent scattering and thermal emission should not change much in such a short interval of wavelength. However CO and $CO_2$ have emission bands in 4.5μm, while 3.6μm is sensitive only to dust (and not $H_2O$, for which the broad emission feature at 2.7 - 3.1 μm barely misses the onset of the 3.6μm pass band). The radial profile of the CO and $CO_2$ gases in the coma is significantly different than that of the dust because the gas species feel photo-dissociative effects as they travel outward from the nucleus. Thus the strength of the emission lines changes as a function of cometocentric distance differently from the dust. This means that the detection and quantification of these species will be straightforward with sufficiently deep imaging of the active comets' comae (Figure 10).

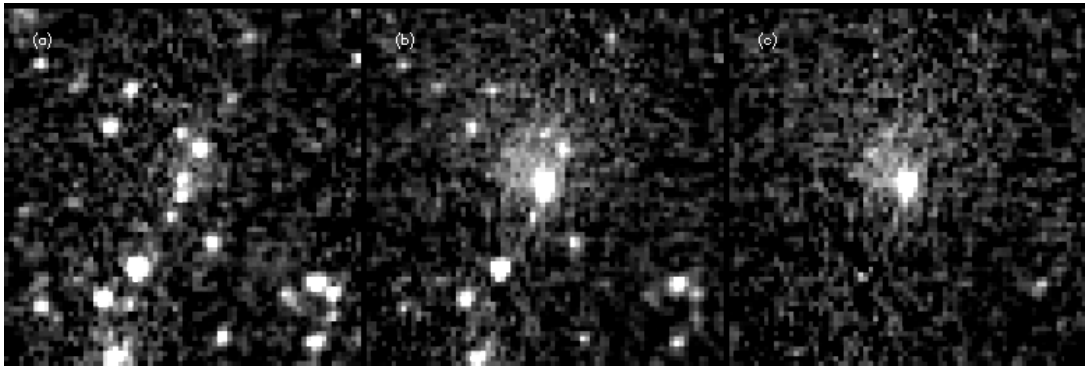

**FIGURE 10.** Comet 2P/Encke as observed in IRAC 3.6μm and 4.5μm in June 2004 by PID 119 (and extracted from SST archive). (A) 88-arcsecond wide subframe of the 3.6μm image centered on the comet. (B) Same as A, but at 4.5μm. (C) Difference of the two images. The coma is dramatically brighter at 4.5μm and it is morphologically distinct.

Furthermore, one can investigate the *interior* of the cometary nuclei by observing the comae at certain critical heliocentric distances. Carbon monoxide sublimates at effectively all distances, but $CO_2$ sublimates only within 13 AU of the Sun, and $H_2O$ does so only within about 3 AU. Furthermore water's phase change from an amorphous to a crystalline phase occurs within 8 AU. The emission of CO and $CO_2$ will change at these heliocentric distances depending on how the various volatile species are mixed within the nucleus.

## 6. ZODIACAL CLOUD

Stars are born surrounded by circumstellar disks of dust and gas. Over timescales of millions of years, dust and gas are removed through radiation forces. However, a few percent of stars older than 400 Myr also have circumstellar dust disks (Habing et al [23]). These older debris disks require sources replenishing system dust over time. Current processes in the Solar System indicate that these sources are comet emissions and asteroidal collisional activity. Stochastic collisional activity can result in large variations in the surface area of dust (Wyatt *et al.* [24]), suggesting that the dust densities and flux levels seen in exo-disks vary strongly with time. This also implies that the Solar System is currently "relaxing" from a time of much denser interplanetary dust created by the collisions that (e.g.) formed the Veritas and Karin asteroid families (Nesvorny et al [25]) -- and with them the zodiacal dust bands -- 5-8 Myr ago. This leads to a key question we wish to address: What does our Solar System's interplanetary dust environment and our zodiacal cloud tell us about the nature of the dusty disks seen around other stars?

The zodiacal cloud results from highly structured dust formations of asteroidal and cometary origin, and is detectable in both thermal emission (Fig. 11) and scattered sunlight (Fig. 12). Both catastrophic collisions (e.g., the formation events for the Karin and Veritas families, that occurred a few Myr ago) and ongoing cometary emission and asteroidal grinding are thought to be source functions for the Interplanetary Dust (IPD). These sources must be continually delivering ~$10^4$ kg/sec of material in order to balance the dust accreted by the Sun and planets, or driven out of the Solar System by solar radiation pressure (Lisse [27]). Exo-zodiacal clouds have been detected by Spitzer, with a frequency of a few percent for mature stars, e.g. the pronounced IR excess found around the 2-10 Gyr K0 star HD69830 (Beichman *et al.* [28]). These exo-clouds are estimated to be 100x or larger the flux emitted by our zodiacal cloud, and thus represent a flux-limited sample, most likely created by recent, stochastic catastrophic asteroid fragmentation events (Lisse *et al.* [1], Wyatt *et al.* [24]).

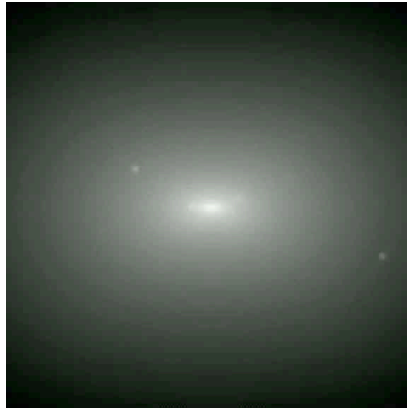

**FIGURE 11.** The Solar System zodiacal dust cloud at 10 μm, as it would appear 1 pc from the Sun with 0.25 AU resolution, 30° above the ecliptic plane. The stretch is logarithmic, and the density profile used for the image is derived from the IRAS model of Good et al. [26]. Jupiter can be seen at the upper left, and Saturn at the lower right.

Solar system zodiacal cloud structures are located from ~0.9 to a few AU, with temperature ranges that make them ideal for mid-infrared detection. Initial characterization of IPD by the COBE DIRBE instrument (Reach, *et al.* [21]) and IRAS (Dermott, *et al.* [29]) has identified the presence of several components, including a smooth cloud, a circumsolar ring, leading and trailing blobs, and dust bands which are associated with asteroid families (e.g., Spiesman *et al.* [30]). Spitzer is uniquely qualified to further characterize the zodiacal background, (e.g. search for dust in resonant orbits around the planets) and since launch, several observing programs have been executed to observe the zody. These projects, however, focus primarily on longer wavelength observations at 24, 70, and 160 μm.

Spitzer's warm mission using the 3.6 and 4.5 *μm* detectors provides an important opportunity to examine the sources contributing to zodiacal background, including the asteroidal dust bands. Reach *et al.* [31], have found that the location of these bands varies with temperature, as observations at different wavelengths sample different parts of the blackbody spectrum. At shorter wavelengths, on the sharply rising Wien side of the SED, warmer dust that lies closer to the sun is detected. The line of sight to this dust appears to be directed further away from the ecliptic plane. At longer wavelengths, such as MIPS 70 and 160 μm, the SED is flat and represents colder, more distant IPD. Observations at these wavelengths are along lines of sight that are directed further away from the sun and appear closer to the ecliptic plane. The above-mentioned Spitzer programs can be supplemented during the warm mission by observing the zody at shorter wavelengths.

Steps in this direction have serendipitously been taken as part of IRAC's nominal calibration plan. During each campaign, IRAC routinely observes the zodiacal background to in the ecliptic plane and at the North Ecliptic Pole to obtain flat fields and sky darks. Figure 12 shows plane and pole data taken during the early part of the mission. The annual variability of the zody is apparent, as Spitzer moves through the circumsolar ring. IPD emission is readily detected, even at the shorter wavelengths.

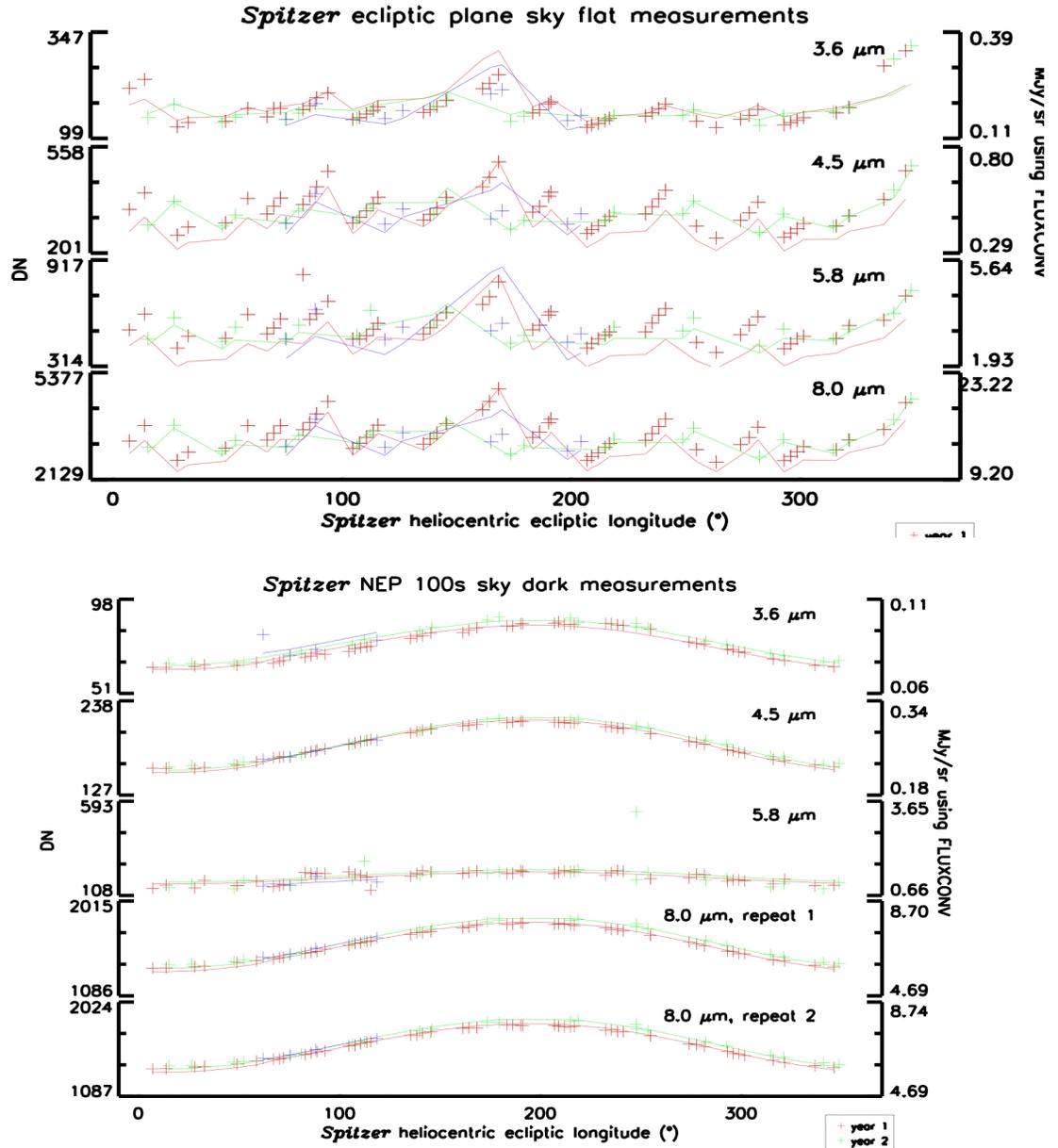

**FIGURE 12**. (Top) North Ecliptic Pole zodiacal background observed by IRAC. Note the annual variability due to tilt of the circumsolar dust ring relative to Spitzer's orbital plane. (Bottom) Ecliptic plane zodiacal background observed by IRAC. Note the biannual variability as Spitzer's line of sight passes through the galactic plane.

During the post-cryo mission, these data that will provide a information on the scale height of the circumsolar ring, as well as other structures cane be enhanced by latitudinal scans that will provide a information on the scale height of the circumsolar ring, as well as other structures.

## 7. ICE GIANTS

Just as the study of the Sun plays critical role in understanding other stars, so do studies of the outer planets in our Solar System yield insight into processes and conditions that occur among the diverse sample of giant planets around other stars. The growing number of extrasolar planetary systems, including many Uranus- and Neptune-sized planets (e.g., Mayor and Queloz [32], Marcy and Butler [33], Rivera *et al.* [34], Beaulieu *et al.* [35], Lovis *et al.* [36]) elevates the importance and significance of understanding the Ice Giants within our own system: Uranus and Neptune. Although most of the giants in exoplanetary systems reside very close to their host stars, it is only a matter of time before such giants are found at larger distances. The results of Warm Spitzer Uranus and Neptune studies will aid studies of extra-solar planets in this size range by elucidating the radiation balance in our local ice-giant atmospheres. Clues to their formation and evolution lie hidden in their chemistry and atmospheric dynamics. We therefore ask the key question: What is the energy budget of the giant planets and what can this tell us about the nature of extrasolar planets?

These are special seasonal times for both Uranus (equinox in 2007) and Neptune (solstice in 2005). HST, Keck, NASA's IRTF, the VLT and many other ground-based observatories have been focused on these planets, often with surprising results as described below. The value of Warm Spitzer data for both planets is that they will provide: (1) a direct comparison with Prime-Mission observations during a time of rapid atmospheric change, and (2) a set of complementary data on two similar yet distinct large bodies, contemporaneous observations with the other seasonal observations.

Theoretical models, grounded primarily in Voyager-era data, are in some cases not adequate for explaining modern observations. What powers their winds? How deep does zonal structure go? What is their atmospheric composition as a function of altitude? How do they interact with incident radiation from the Sun? Of particular interest are the relative roles of dynamics and insolation in controlling atmospheric properties, along with the timescales and phase lags associated with significant seasonal change.

### 7.1 Uranus

Obliquity plays an important role in climate change on Earth (e.g., Zachos *et al.* [37]) and Mars (e.g., Nakamura and Tajiki [38]). Within the field of planetary atmospheres, Uranus' atmosphere is uniquely important because of its extreme obliquity (98 deg). The markedly low value of its internal heat source relative to each of the other giant planets also enhances its seasonal extremes from both radiative and dynamical perspectives.

During—and in the decade following—the 1986 Voyager flyby, when Uranus was near solstice, its atmosphere was quiescent and nearly featureless. Most atmospheric models of Uranus are based from data taken in this epoch or earlier. In late 2007, a rare opportunity is permitting us to study Uranus in detail: the Earth and the Sun are crossing the planet's equatorial plane (Fig. 13). This geometry takes place just twice every Uranus year, i.e., every 42 Earth years.

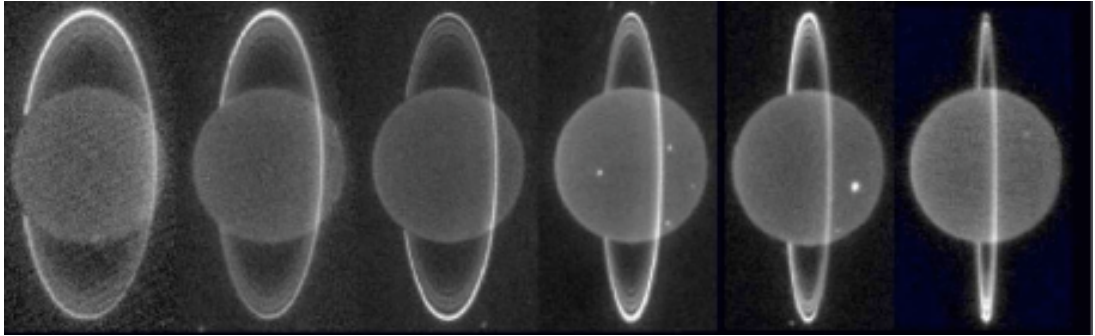

**FIGURE 13.** Annual Keck images of Uranus from 2001 to 2006 showing the impending 2007 equinox. This time sequence highlights the changing viewing aspect as Uranus approaches the equinox, or Ring-Plane crossing (RPX). The images were obtained with adaptive optics at 2.2 μm, where methane and hydrogen absorb sunlight in the planet's atmosphere. The white spots in the last three years are high-altitude cloud activity, possibly triggered by seasonal insolation changes. [Images from de Pater et al. [39], updated with 2006 data by HBH.]

Some Uranus atmospheric models predicted a radiative seasonal response (e.g., Wallace [40]; Friedson and Ingersoll [41]). However, those models predicted that the planet's long radiative time constant at all levels below the upper stratosphere would permit only small physical seasonal changes within the visible atmosphere, with phase lags of order 10-20 years (Friedson and Ingersoll [41]). In contrast, observers have been seeing notable atmospheric change on Uranus as equinox approaches, apparently with no phase lag. Hubble Space Telescope (HST) images of Uranus showed the south polar region darkened as it began to receive less direct sunlight (Rages *et al.* [42]). Pre-equinox images from HST and from the Keck 10-meter telescope in Hawaii (Hammel *et al.* [43], [44]; Sromovsky and Fry [45]) showed tremendous tropospheric cloud activity across the planet (Fig. 14), including changes on time-scales as short as days (Hammel *et al.* [43]).

At equinox, solar forcing is symmetric over the two hemispheres. Hemispheric asymmetries at equinox may be a measure of a seasonal radiative phase lag or may alternatively be the signature of a much more rapidly evolving dynamical instability forced by the reversal of insolation between warm and cold hemispheres. At present, the two hemispheres differ strongly in the albedo of tropospheric cloud bands and in the dynamical activity producing localized bright and dark cloud features (Fig. 14).

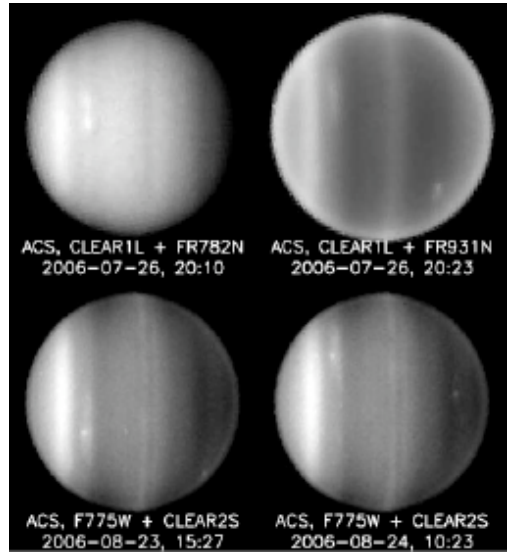

**FIGURE 14.** Uranus images from HST. These images from the Advanced Camera for Surveys on the Hubble Space Telescope illustrate the significant asymmetry in the troposphere of Uranus. [Images courtesy L. Sromovsky (University of Wisconsin).]

One manifestation of enhanced dynamical activity could be increased generation of upwardly propagating gravity waves and hence an increased eddy diffusion coefficient in the stratosphere. Methane would get carried up to higher altitudes, and photochemical products like acetylene and ethane would be produced at a higher rate. As a consequence, hydrocarbon species may be more readily detectable. Encrenaz *et al.* [46], using 1996 observations with the Infrared Space Observatory (ISO), derived an acetylene abundance on Uranus that was 30 times greater than was seen during the 1986 Voyager encounter (Bishop *et al.* [47]). The high-SNR disk-averaged Spitzer IRS spectra of Uranus (Fig. 15) confirm the detection of ethane, as well as other hydrocarbon species. More importantly, the spectra also show evidence for seasonal, or at least temporal, change between ground-based data in the Voyager era and the newer Spitzer data.

Realistic models of the atmosphere of Uranus are needed to explain the plethora of time-variable modern observations described above. Data from the Warm Spitzer era - together with additional data from the Spitzer Prime Mission as well as SMA, VLA, VLT, and many other facilities - can be used to create realistic physical models of the planet's changing infrared emission. The 3.6- and 4.5-μm data taken during the Warm Era can be directly compared with Prime-Mission measurements at those wavelengths. Since the atmosphere of Uranus is demonstrably changing right now, a cadence of observations of order once a year, or once every six months, can capture time-scales of change.

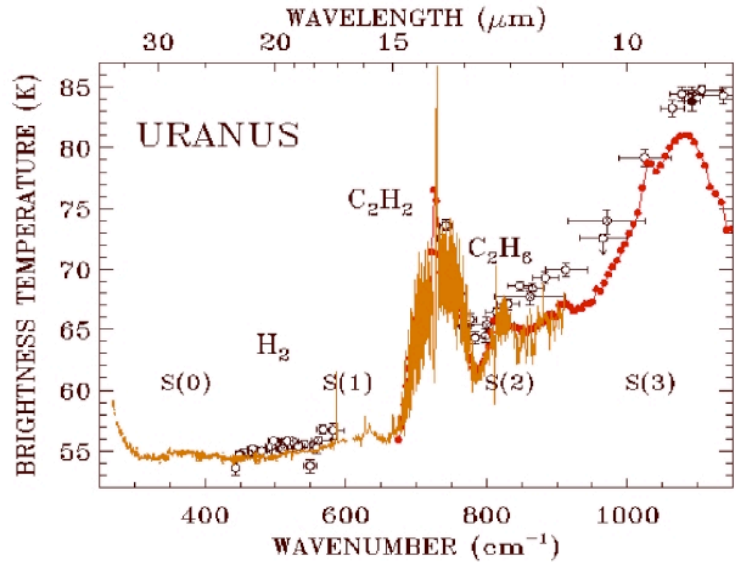

**FIGURE 15.** - Uranus changes as detected by Spitzer. This proprietary figure, provided by G. Orton, illustrates the change in the mid-infrared spectrum of Uranus during the past two decades (Orton et al. [48]). The open symbols are ground-based mid-infrared data from the mid-1980s, compared with current-era Spitzer data (lines and small closed symbols). Currently Uranus has lower brightness temperatures throughout this spectral region.

## 7.2 Neptune

The more distant ice giant, Neptune, has also been exhibiting significant atmospheric change in recent decades, though the driving mechanisms are less well understood that for seasonally-dominated Uranus.

At optical wavelengths, Neptune in 2003 reached its brightest level in nearly 30 years of photometric monitoring (Lockwood and Jerzykiewicz [49]; Hammel and Lockwood [50]). The planet has exhibited pronounced cloud activity throughout this time, possibly undergoing a transformation from one cloud distribution to a new configuration (Hammel and Lockwood [50] and references therein). Hammel and Lockwood [50] predicted that the stratospheric temperature may have increased on Neptune since the mid 1980s, because stratospheric temperatures on Uranus (Young et al. [51]) rose and fell in lockstep with that planet's visible wavelength reflectivity (Hammel and Lockwood [49]).

In the mid-infrared, early 7-13 µm IRTF observations of Neptune (Orton *et al.* [52], [53]; Hammel et al. [54]) hinted at variation in the 12.2-µm ethane emission feature. Data taken a decade later showed significantly increased ethane and methane emission (Hammel *et al.* [55]). Radiative-transfer models, discussed in detail in Hammel *et al.* [55], suggested that increasing stratospheric temperature has been driving the long-term mid-infrared change, an interpretation supported by independent observations at sub-millimeter wavelengths (Marten et al. [56]) and also by the visible-wavelength photometry discussed above. Recent images of Neptune at methane (7.7-um) and ethane (12-um) emission bands with the MICHELLE instrument on Gemini (Hammel *et al.* [57]) showed enhanced south polar emission (Fig. 16, upper panel), similar to that of Saturn (Orton and Yanamandra-Fisher [58]). Both enhancements probably result from long periods of continuous exposure to sunlight (i.e., are seasonal in nature).

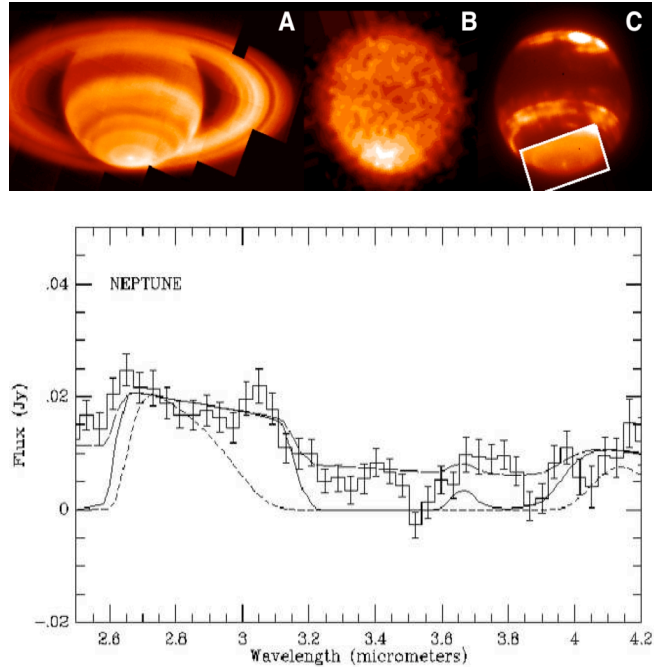

**FIGURE 16.** Neptune images and spectra. (Upper panel) Neptune compared with Saturn (Hammel et al. [57]). Image A shows Saturn at 8.0 µm (Orton and Yanamandra-Fisher [58]) with strong methane emission from its southern pole. Image B shows Neptune at 7.7 µm taken with the Gemini telescope; as with Saturn, the methane emission arises from the south polar region. (Note that Saturn and Neptune are not displayed to the same scale.) In Image C, Neptune -- taken at 1.6 µm with the Keck Adaptive Optics system -- reveals zonal circulation that is as tightly confined in the polar region as that of Saturn. (Lower panel) Neptune spectrum from 2.5 to 4.2 µm (Encrenaz et al. [59]). The ISOPHOT-S spectrum (data with error bars) is compared with atmospheric models (lines) using the parameters of Fink and Larson [60] with differing column densities of methane. Most other wavelengths exhibit temporal variation (see text), thus Warm Spitzer observations will likely also differ from earlier Spitzer and ISO observations. Comparisons of Warm Spitzer with data and models as shown here will permit methane column densities to be determined as a function of time.

The Warm Spitzer IRAC data are sensitive to intermediate wavelengths, and likely also bear evidence for atmospheric change, since these regions are acutely sensitive to methane (Fig. 16, lower panel). The most important contribution Warm Spitzer will make for Neptune is in the comparison with Prime-Mission observations, and in obtaining adequate temporal samples to characterize the time-scales of atmospheric change. Yearly samples are a minimum sample rate; twice per year would provide a more robust sampling since short-term variations are seen at wavelengths ranging from the optical to the mid-infrared (Hammel et al. [54], [55]).

# 8. OBSERVING TIME ESTIMATES AND STRATEGIES

## 8.1 Small Point Sources

The limiting fluxes for the small body point-source programs are all estimated in a very similar manner, so they are summarized together here. The limiting sensitivity for each project/group of objects is constrained by the desired S/N and the longest desired integration time. For illustrating potential observing time to perform key Solar System observations, we have constrained the longest AOR times to ~1 and 2 hours. The required S/N for each group of objects is listed in the table below, and summarized in the individual science sections above. We have assumed the Warm Era sensitivities for IRAC 3.6μm and 4.5μm are the same as in the Cold Era.

Reflected fluxes in IRAC 3.6μm and 4.5μm are estimated from known visible magnitudes. For the 30 TNOs, Centaurs, and Trojans observed in program 20769, the mean color indices are V-3.6μm ~ V-4.5μm = 1.6. For the calculations here, we used a slightly more pessimistic estimate of V-3.6μm = V-4.5μm = 1.5. Visible fluxes were calculated (using the IAU Minor Planet Center ephemeris generator) for all TNOs and Centaurs at three month intervals for the first three years of the Spitzer warm mission (March 2009 - March 2012). The above color indices were then applied to estimate the fluxes in the IRAC channels of each object.

The table below summarizes the estimated sensitivities expected for each channel and the corresponding number of observable objects (i.e. those with fluxes greater than or equal to the limiting sensitivities). The 3σ confusion limit in clean, low-background fields is ~1.8 μJy. Because we have little choice of where to observe our targets, we must assume that confusion will be significantly worse than 1.8 μJy. The magnitude distribution is a power-law, with many more faint objects than bright. As a simplification, we assume below that all observable targets will benefit from being observed twice at slightly different epochs, with the image pairs being used to subtract away confusing background sources (see *e.g.* Stansberry *et al.* [61]). The sky-subtracted image pairs can also be shifted and co-added to increase the SNR of the detections by about 1.4.

For observations of reflected sunlight (all objects except NEOs), 4.5μm provides the more stringent constraint on observability. For detecting thermal flux from NEOs, 3.6μm provides the more stringent constraint. Almost all presently known PHAs have maximum IRAC 3.6μm/4.5μm fluxes during the Spitzer era greater than 1 μJy. Furthermore, around half of all known PHAs have maximum IRAC 3.6μm /4.5μm fluxes during the Spitzer era greater than 1 mJy. Consequently, observing these objects should be quite easy. (The only complication is that, because these objects have orbits ~1 AU, their fluxes as observed by Spitzer vary by enormous factors during the Spitzer era as the Sun-asteroid-Spitzer geometry evolves. These factors are calculable, and have been taken into consideration for the technical summary above.)

*A base level program for each type of object would observe, using both channels, at least 100 - 200 objects that are brighter than the limiting flux. Many objects are significantly brighter than the limiting flux, so would require shorter observing times.*

We have grouped objects into coarse brightness bins to estimate the total time required to observe all the observable objects.

**TABLE 1. Observeable Objects**

| Objects | Channel | S/N | Sensitivity µJy | $T_{exp}$ seconds | dithers | AOR time | # of Potential Targets |
|---|---|---|---|---|---|---|---|
| TNO/Centaurs | 2 | 3 | 1.1 | 100 | 32 | ~1hr | 140 |
| TNO/Centaurs | 2 | 3 | .8 | 100 | 64 | ~2h | 213 |
| Trojans | 2 | 10 | 3.7 | 100 | 32 | ~1hr | 1436 |
| Trojans | 2 | 10 | 2.6 | 100 | 64 | ~2hr | 1694 |
| Main Belt | 2 | 10 | 3.7 | 100 | 32 | ~1hr | ~10,000 |
| Main Belt | 2 | 10 | 2.6 | 100 | 64 | ~2hr | ~10,000 |
| NEOs | 1 | 10 | 3.7 | 100 | 32 | ~1hr | ~2000 |
| NEOs | 1 | 10 | 2.6 | 100 | 64 | ~2hr | ~2000 |
| Inactive comets | 1 | 10 | 3.7 | 100 | 32 | ~1hr | ~100 |
| Inactive comets | 2 | 10 | 2.6 | 100 | 64 | ~2 hr | ~150 |
| Active comets | 1 | 10 | 3.7 | 100 | 32 | Variable | ~500 |
| Active comets | 2 | 10 | 2.6 | 100 | 64 | Variable | ~500 |

## 8.2 KBO/Centaur/Trojan/Outer MBA Observing Strategy

One potential strategy is to make two observations of each object, separated by enough time for the object to move at least 30", but remain in the 5.2' x 5.2' FOV. There are four advantages to this strategy: (1) certain identification of the object by its motion in the field, (2) background subtraction (this technique is effectively a shadow observation, alleviating the negative effects of confusion/high background near the ecliptic, in which we collect data in both frames rather than just sky in one), (3) if our moving object happens to pass in front of a background source in one observations, we will still collect data from the other (this avoids very detailed timing constraints), 4) for objects with known rotation periods, the second observations would be phased with the object's rotation so that opposite hemispheres are observed in order to search for rotational variability. Dithered observations will allow removal of cosmic rays and stray light correction. Assuming the flux from these objects is spread over four pixels, then fluxes of some of them are comparable to the "high" background flux. The follow-on "shadow" observations are therefore crucial and will allow accurate removal of the background.

## 8.3 Comet Trail Observing Strategies

Figure 17 below presents estimates of comet trail surface brightness assuming $\sim 5 \times 10^{-9}$ and particle albedos of 0.05 for the scattered light component. Within 2 AU, with some effort, trails will be detectable in IRAC 3.6µm and/or 4.5µm. More than 140 comets come within this distance to the sun and some fraction of them will be available for study during the Spitzer Warm Cycle. Using SENS-PET integration times of less than an hour allows for 1s detection of 0.002 and 0.003 MJy/sr in IRAC 3.6µm and 4.5µm, respectively, with a medium background. Because trails are extended in their orbital direction, increase S/N can be obtained by integrating in that direction.

Eight comet dust trails were detected by IRAS. This was expanded to more than 30 trails during the Spitzer cryo-phase. This may be more than tripled during the Spitzer warm phase.

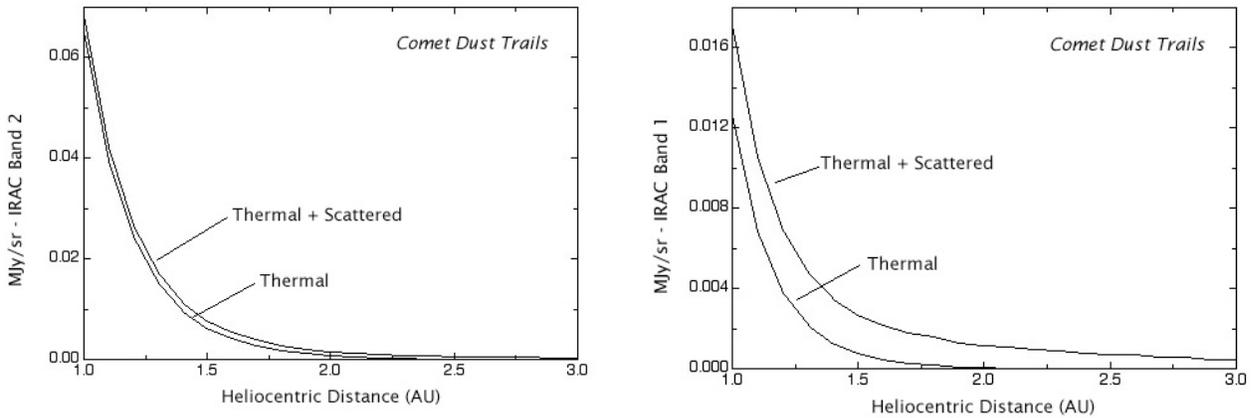

**FIGURE 17.** Trail surface brightness in the functioning IRAC bands during the Spitzer Warm Phase are shown assuming a near-infrared reflectivity of 0.05 and color temperature of 310K at 1 AU (Sykes et al. [19]) and median optical depth of $5 \times 10^{-9}$ (After Reach et al. [21]).

## 8.4 Observations of Active Comets

Predicting surface brightnesses of active comets is famously difficult (cf. Kohoutek in 1973), but we can make a rough estimate. We note that in the Cold Era, for 3200 seconds of integration (100 second frame time and 32 dithers), one can achieve S/N=10 in a pixel in 3.6µm with a surface brightness of 0.97 µJy/pixel and in 4.5µm with 1.22 µJy/pixel. A typical case is the IRAC observation of comet 2P/Encke by Gehrz (PID 119) in June 2004 when the comet was 2.6 AU from the Sun. In only 120 seconds of integration at 3.6µm and 4.5 µm, the S/N in the inner coma was sufficiently high to detect differences in the coma morphology (as described in the Carbonaceous Species in Comets section). Comet Encke has a "typical" infrared coma and so we can conservatively estimate that ~50% of the ~350 Short Period comets will have observable comae when they are within 3 AU of the Sun. Thus at the very minimum it will be possible to perform a survey for $CO/CO_2$ with a sample encompassing a large fraction of the known population.

Predicting the activity of comets out to 13 AU is more difficult however, since observations of such comets have so far been scarce. As a point of comparison, we can use some of the preliminary results obtained by one of us (YRF) in a current Spitzer survey of "distant" (4-5 AU) Short Period comets (PID 30908). Approximately 15% of the sample shows activity, revealing itself as thermal emission of dust in either the IRS PU bands or the MIPS 24 µm band. This gives us confidence that there are a significant number of Short Period comets that are active beyond the nominal water-sublimation distance. Naturally the fraction of comets that are active diminishes as heliocentric distance increases, but we can use the Cold Era surveys and observations as guides to specifically target comets known to be active while far from the Sun.

## 9. REQUIRED SSC RESOURCES AND TOTAL SPITZER WARM ERA OBSERVING TIME

Observing all available objects is not necessary nor advocated to address the key questions identified in this proposal. The times listed in the above table represent the maximum time required to exhaustively survey the identified targets. In practice, many subsets of these times will likely be requested to address a range of topics, ranging from investigations of individual objects, to studies of small groups, and on up to larger surveys.

*Overall, the application of Spitzer's IRAC 3.6*µm *and 4.5* µm *in the warm era to the important planetary science questions presented here will require a number of resources to be in place at the SSC. These include the following:*

- A large, current database of Solar System orbital elements.
- Staff members knowledgeable in moving target observations.
- The ability to ingest similar AORs for a large list of diverse objects.
- ~1000 hours of observing time with IRAC 3.6 and 4.5 µm to perform the useful science presented herein.